\documentclass[aps,prd,twocolumn,showpacs,groupedaddress,floatfix]{revtex4}

\usepackage{graphicx}
\usepackage{longtable}
\DeclareGraphicsExtensions{.pdf, .eps, .jpg, .png}

\newcommand{\dee}{\partial}
\newcommand{\beam}{\ensuremath{\theta_\mathrm{fwhm}}}
\newcommand{\fsky}{\ensuremath{f_\mathrm{sky}}}

\begin{document}


\title{Next-generation test of cosmic inflation}



\author{Benjamin Gold}
\email{gold@bubba.ucdavis.edu}
\author{Andreas Albrecht}
\affiliation{Department of Physics, UC Davis,
One Shields Avenue, Davis CA, USA 95616}


\date{\today}

\begin{abstract}
The increasing precision of cosmological datasets is opening up new
opportunities to test predictions from cosmic inflation.  Here we
study the impact of high precision constraints on the primordial power
spectrum and show how a new generation of observations can provide
impressive new tests of the slow-roll inflation paradigm, as
well as produce significant discriminating power among different
slow-roll models.  In particular, we consider next-generation
measurements of the Cosmic Microwave Background (CMB) temperature
anisotropies and (especially) polarization, as well as new
Lyman-$\alpha$ measurements that could become practical in the near
future. We emphasize relationships between the slope of the power
spectrum and its first derivative that are nearly universal among
existing slow-roll inflationary models, and show how these
relationships can be tested on several scales with new observations.
Among other things, our results give additional motivation for an
all-out effort to  measure CMB polarization.  
\end{abstract}


\maketitle


\section{Introduction}

Over the last several years, extraordinary observational support has
emerged for the idea that key features of our universe were formed by a
period of cosmic inflation.  During inflation, the universe enters a
period of ``superluminal expansion'' which imprints certain features
on the universe.  The physical degree of freedom responsible for
inflation, generically called the ``inflaton'', has yet to find a
comfortable home in fundamental theory, and there are many competing
ideas for how fundamental aspects of inflation could play out.  None
the less, at the phenomenological level a standard picture of inflation
has emerged.  

From the observational point of view, the standard picture
is defined by a set of observable characteristics that are the same
across virtually all proposed models for the inflaton.  The most well
known predictions from the standard picture of inflation are that
the universe 
has critical density (to within roughly one part in $10^5$), that the
primordial perturbations are coherent, (leading, for example, to acoustic
peaks in the microwave background power spectrum), and that the power
spectrum of primordial perturbations is nearly scale invariant, with
the tilt parameter $n_s$ constrained to be close to unity.  A unique
spectrum of coherent gravitational waves is also predicted, which
could eventually come within range of direct gravitational wave
detectors, and which could also be observed indirectly via signals in
the microwave background polarization.

But inflation makes many more predictions than these.  Specifically, a
given model for the inflaton will predict a detailed shape for the
primordial power spectrum that goes way beyond what can be described
simply by a single tilt parameter.  The detailed shape of the power
spectrum is a reflection of the particular evolution of the inflaton
during inflation, something that is precisely specified in a given
model.  In this paper we show how the next generation of experiments
could bring studies of the power spectrum shape to a whole new
level.  These studies present two kinds of opportunities:  One
opportunity is to make additional tests of the standard picture of
inflation.  To this end, we focus on particular power spectrum
features that are known to exist across essentially all 
inflation models.  The search for these features could either confirm
or falsify the standard picture of inflation.  

The second opportunity is to go beyond broad tests of the standard
picture.  The next generation of experiments which we consider here will
provide important additional information.  This information could
actually distinguish among different specific inflaton models,
assuming the standard model is not falsified, or it could provide very
useful constraints on the alternatives if the standard model is ruled
out. 

Our approach here is similar to and inspired by a large body of 
earlier work on this subject
\cite{ckll,lk,kt,jkks,cgl,dkk,kmr,hann1,hk}.
Our emphasis here is identifying what useful information about the
primordial power spectrum and inflation might be revealed by a new
generation of experiments. For this work we assume that the primordial
perturbations are adiabatic.  As emphasized in \cite{Bucher:2000cd},
relaxing this assumption would result in more degeneracies and would
lead to somewhat weaker constraints on parameters.  

The organization of this paper is as follows: Section \ref{sec:inflation} gives
background information about slow roll inflation.  Section
\ref{ssec:slow roll} introduces the aspects of slow roll inflation we intend to test.
Section \ref{sec:experiments} discusses the CMB and Lyman-$\alpha$
data (existing and 
simulated) we use to test inflation.  Section \ref{sec:results} gives our main
results and \ref{sec:conclusion} gives our conclusions.   Appendix \ref{appendixA}
gives details of the inflation models we use for our plots.


\section{Scalar Field Inflation}
\label{sec:inflation}

In the standard picture, inflation occurs when the
potential energy density $V(\phi)$ of a scalar field $\phi$
(the inflaton) dominates the stress-energy
\cite{guth,as,linde,lindechaos}.  This scalar field may be a
true scalar field or an effective field obtained from some more
complicated theory.  The period of 
potential domination is usually closely connected to very slow
evolution of the inflaton field, the so-called ``slow roll'' behavior,
and it is this slow evolution that produces a nearly scale invariant
spectrum of perturbations.  In the slow roll inflationary scenario,
however, $\phi$, and therefore $V(\phi)$, are not 
completely constant during inflation, and this leads to deviations
from total scale invariance.   

The dynamic behavior of the potential is determined by the equation of motion for a scalar field in an expanding universe
\begin{equation}
	\ddot\phi + 3 H \dot \phi + V'(\phi) = 0 .
\label{Eqn:phiddot}
\end{equation} 
The gradient of the field is ignored, as even if present it will be
quickly damped by the inflationary expansion to the degree that it is
irrelevant for the classical evolution of the background spacetime, which is
what we determine from Eqn. \ref{Eqn:phiddot}.  The field is considered
to be in a slow roll regime if $\ddot\phi$ is negligible. 
The Hubble constant $H \equiv \dot a/a$ is related to the total energy
density of the universe which if dominated by the scalar field is 
\begin{equation}
	H^2 = \frac13 \left( \frac12 \dot\phi^2 + V(\phi) \right), 
\end{equation}
where $M_P \equiv 1/\sqrt{8\pi G}$ has been set to unity.  
It is customary to define slow roll parameters such as
\begin{equation}
	\epsilon \equiv \frac12 \left( \frac{V'}V \right)^2 , \quad
	\eta \equiv \frac{V''}V , \quad
	\xi^2 \equiv \frac{V'V'''}{V^2} ,
\end{equation}
although several other conventions also exist in the literature.
Assuming that these parameters (and the higher derivatives of $V$) are
small leads to expressions for the primordial amplitude of density
perturbations \cite{bard} 
\begin{equation}
	|A_S(k)| = \frac{\delta\rho}\rho = \frac{H}{\sqrt{2\epsilon}} ,
\end{equation}
the spectral index of these perturbations
\begin{equation}
	n_S(k) \equiv \frac{d \ln k {A_S}^2(k)}{d \ln k} = 1 + 2\eta -
	6\epsilon, 
\end{equation}
and the derivative of the spectral index
\begin{equation}
	{n_S}'(k) \equiv \frac{d n_S(k)}{d \ln k} = -2\xi^2 + 16 \eta
	\epsilon - 24 	\epsilon^2. 
\end{equation}
The right-hand side of each equation above is to be evaluated during
inflation at the time when the scale of interest $k$ exits the Hubble radius.

\subsection{Slow roll, $n_S$, and ${n_S}'$}
\label{ssec:slow roll}

A general feature of slow roll is that ${n_S}'$ is higher order in the
slow roll parameters than $n_S$.  Thus if we assume higher order terms
become increasingly small, then barring a conspiracy of cancellation between
terms $|{n_S}'| \sim (n_S-1)^2$ or less.   
As pointed out in \cite{stew1}, while this assumption is commonly made
it is an addition to the common assumption that the slow roll
parameters are small, at least when formally considering ``the space
of all possible inflation models''.
We emphasize here that in practice the slow roll hierarchy between
$n_s$ and $(n_s)'$ is indeed realized in the vast majority of
published models, so detection of a large ${n_S}'$, while not completely
ruling out slow roll, would force a rethinking of the standard picture of
inflation.
This relation can be generalized to higher derivatives, resulting in a kind
of consistency relation
\begin{equation}
	\label{eq:dconsistency}
	\left|\frac{d^n n_S}{d \ln k^n}\right| \leq |n_S-1|^{n+1} , 
\end{equation}
which could be taken as defining a kind of `normal' class of inflationary models.
\begin{figure}
	\includegraphics[width=3.4in]{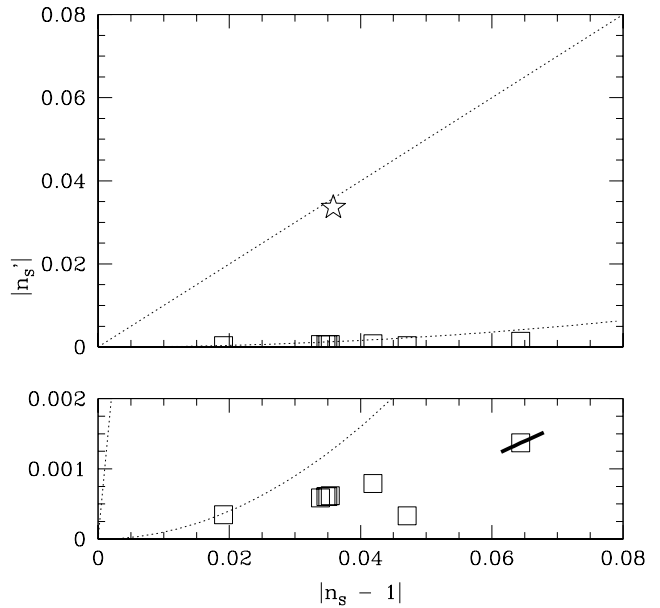}
	\caption{
		\label{fig:models}
		The values of $n_S$ and ${n_S}'$ at CMB scales are plotted for several models.
		The lower graph is a zoomed view of the bottom of the upper graph. In both,
		the upper line is ${n_S}' = n_S - 1$ and the lower line is ${n_S}'
		 = (n_S-1)^2$. The thick line on the lower graph shows the evolution of $n_s$
		and ${n_S}'$ with scale for two orders of magnitude in each direction. 
	}
\end{figure}

For inflationary scenarios involving multiple fields (often called
hybrid models) this condition is relaxed.  With multiple fields the
extra freedom introduced makes it  easier to remain on the edge of
violating the slow roll conditions over many e-foldings.  Thus the
tendency for models to lie under the curve $|{n_S}'| = (n_S-1)^2$
is not as strong for hybrid models.
Of course, theories which generate primordial density perturbations from something
other than inflation also have no need to obey the above consistency relation.

To discuss observational constraints, we take the point of view of \cite{tz}
that the primordial power spectrum is an 
unknown function, which may be sampled by experiments
at one or more scales.  Statements about the slope of this function
(and higher derivatives) then can only be tested by effectively sampling the
function to high accuracy at several nearby scales.  Current analysis
tends to use all the data to provide only limited information
about the power spectrum.  We wish to emphasize that 
higher quality data over a range of scales will allow us extract
significantly more information about the primordial power spectrum,
information that can have a great impact on tests of the inflationary picture.

\subsection{Model space}

One of the simplest models to evaluate is a pure
exponential.  For $V=V_0 \exp (\lambda \phi)$ the spectral index $n_S
= 1 - \lambda^2$ for all scales, and thus ${n_S}'$ and all higher
derivatives are zero.  Thus for this model measurements of $n_S$
simply map into constraints on $\lambda$, without presenting an
opportunity to falsify the general model.  But a measurement of
${n_S}'$ which excludes zero can rule this type of inflaton potential
altogether. 

This model is special in that the potential is constructed to
form the simplest possible power law spectrum of perturbations.  Most
inflationary models have 
more complicated forms, but many proposed models approximate  the
exponential behavior on the scales which cosmological measurements
probe.  

A previous survey of models of inflation and their spectral index can be found in \cite{lyth}.
In figure \ref{fig:models} several different models taken from a
sampling of the literature have been plotted. (Specific information
about each model plotted is given in the Appendix.)  The types of models
range from high-order polynomial to mass-term to brane-world inspired
scenarios \cite{ll,dt,guthpi,brane1,lindechaos}.  Despite the
difference in the form of each model's potential, almost all of these live on
or below the line $|{n_S}'| = (n_S-1)^2$. 

Certain models can exhibit more exotic behavior, such as the running-mass model described in \cite{clm} (an example this type of model is marked by the star on Figure 1), or the interesting type of potentials given by
Stewart and Lyth\cite{stew1}. These can give $|{n_S}'| \sim |n_S-1|$ over a
range of scales, resulting in a
markedly different primordial power spectrum.
These models form an important `alternate' class of models which will be easy for future data to confirm or rule out.

\section{Determining how well experiments can do}
\label{sec:experiments}

\begin{figure}
	\includegraphics[width=3.4in]{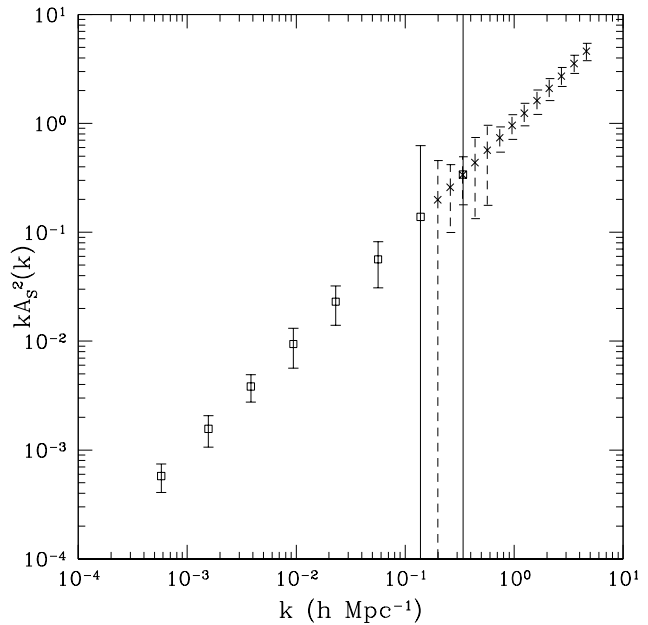}
	\caption{
		\label{fig:current}
		One-sigma error in the (binned) primordial power
		spectrum from a WMAP-like microwave 
		anisotropy
		temperature and polarization experiment (squares with solid error bars),
		and from a Lyman-$\alpha$ experiment as in \cite{croft} (crosses with
		dashed error bars).
	}
\end{figure}

\begin{figure}
	\includegraphics[width=3.4in]{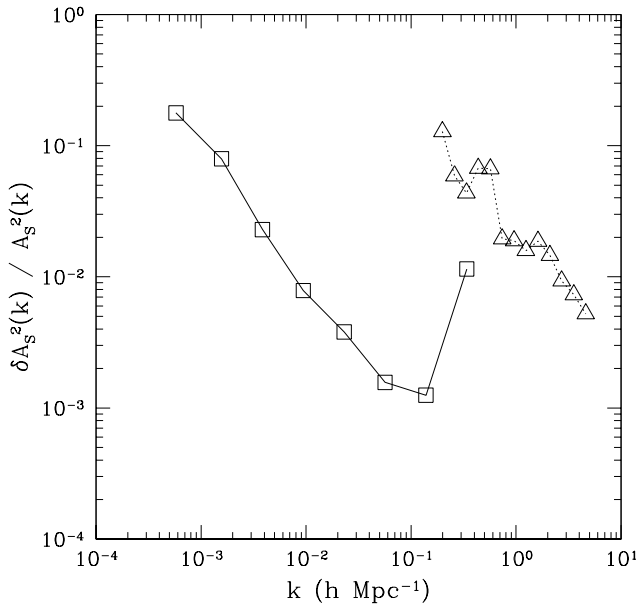}
	\caption{
		\label{fig:super}
		One-sigma error in the (binned) primordial power spectrum from an
		after-Planck temperature
		and polarization experiment (squares), and from a Lyman-$\alpha$
		experiment with total uncertainties one hundred times
		smaller than current experiments (triangles).
	}
\end{figure}

To find the possible impact of CMB and Lyman-$\alpha$ experiments, we model
the primordial power spectrum with a Taylor series expansion of the spectral
index around a particular scale
\begin{equation}
	{A_S}^2(k) = P\left(\frac{k}{k^*}\right)^{-1+n_S+{n_S}' x + \frac12 {n_S}'' x^2 +
	\ldots},
\end{equation}
where $x \equiv \ln(k/k^*)$, $k^*$ is the pivot point, and $P$ is an overall
normalization.  We then use Fisher matrix techniques to jointly estimate parameters
for each experiment.

Parameterizing the power spectrum a particular way has its own set of
advantages and disadvantages.  One nice feature of the step-wise
form of \cite{wss} is that it is easy for each bin amplitude to be a nearly statistically independent parameter in the likelihood analysis.  The coefficients of
a Taylor series expansion generally have larger covariances.  The
disadvantage of the step-wise form, however, is that the quantities of interest to us 
(such as the spectral index) are not simply related to the shape parameters.

\subsection{\label{ssec:CMB} CMB}

We first consider CMB experiments which measure both the temperature and polarization anisotropies.  For scalar perturbations there are three power spectra described by $C_\ell^{(T,E,C)}$, where $(T,E,C)$ indicate the temperature, $E$-mode polarization, and cross-correlation power spectra.  These $C_\ell$ all have similar dependence on the primordial power spectrum, and are found by

\begin{equation}
	C_\ell = (4 \pi^2) \int \frac{dk}{k} {A_S}^2(k) \left| \Delta_\ell(k, \tau=\tau_0) 	\right|^2 , 
\end{equation}
where $\Delta_\ell(k, \tau=\tau_0)$ are transfer functions
for the CMB, and ${A_S}^2(k)$ is the squared amplitude of
the primordial power spectrum.  The functions $\Delta_\ell(k,
\tau=\tau_0)$ depend on 
cosmological parameters, and can be conveniently calculated using
the CMBFAST code \cite{CMBFAST}. 

Error in a cosmological parameter $s_i$ can be estimated as $\sqrt{(\mathbf{F}^{-1})_{ii}}$, where $\mathbf{F}$ is the Fisher matrix
\begin{equation}
	\label{eq:fisher}
	{F}_{ij} = \sum_{I,J = (T,E,C)}\sum_\ell \frac{\dee C^I_\ell}{\dee s_i}
	(\mathbf{C}^{-1})_{IJ} \frac{\dee C^J_\ell}{\dee s_j}.
\end{equation}
The covariance matrix $\mathbf{C}$ has elements which can be approximated by \cite{lk,sz_pol}
\begin{eqnarray} \label{eq:cov_begin}
	\mathrm{C}_{TT} & = & \frac{2}{(2\ell + 1) \fsky} 
	\left( C^T_\ell + N^T_\ell \right)^2 ,
	\\
	\mathrm{C}_{EE} & = & \frac{2}{(2\ell + 1) \fsky} 
	\left( C^E_\ell + N^E_\ell \right)^2 ,
	\\
	\mathrm{C}_{CC} & = & \frac{1}{(2\ell + 1) \fsky} 
	\nonumber \\ && \times
	\left[({C^C_\ell})^2 + 
	\left( C^T_\ell + N^T_\ell \right)
	\left( C^E_\ell + N^E_\ell \right) \right] , 
\end{eqnarray}
for the diagonal elements, and the off-diagonal elements are
\begin{eqnarray}
	\mathrm{C}_{TE} & = & \frac{2}{(2\ell + 1) \fsky} 
	({C^C_\ell})^2 , 
	\\
	\mathrm{C}_{TC} & = & \frac{2}{(2\ell + 1) \fsky} 
	C^C_\ell
	\left( C^T_\ell + N^T_\ell \right),
	\\
	\mathrm{C}_{EC} & = & \frac{2}{(2\ell + 1) \fsky} 
	C^C_\ell
	\left( C^E_\ell + N^E_\ell \right). \label{eq:cov_end}
\end{eqnarray}
In equations \ref{eq:cov_begin} through \ref{eq:cov_end}, $\fsky$ is the fractional sky coverage of the experiment, and we have defined a noise term 
\begin{equation}
	N^{(T,E)}_\ell \equiv \sigma^2_{(T,E)} \beam^2
	e^{(0.425 \, \beam \ell)^2} , 
\end{equation}
where $\sigma_{(T,E)}$ is the noise per pixel in the temperature and polarization measurements and $\beam$ is the width of the beam.  For experiments like WMAP which obtain temperature and polarization data by adding and differencing two polarization states, the noise per pixel for each is related by ${\sigma_T}^2 = {\sigma_E}^2 / 2$.

The derivatives $\dee C_\ell / \dee s_i$ are evaluated via finite difference using a numerical code derived from CMBFAST and DASh \cite{dash}.  We consider only flat models, using as parameters
\footnote{
For an analytic expression for $\ell_A$, see \cite{hu02}.  These cosmological parameters correspond closely the the $\cal A, \cal B, \cal M,$ and $\cal Z$ parameters of \cite{kmj}.
}
the acoustic angular scale $\ell_A$, $\Omega_b h^2$, $\Omega_m h^2$, $\exp(-2\tau)$, the primordial power spectrum normalization $P$, and the first seven coefficients in the expansion of $n_S$.
We use $0.05 h\textrm{ Mpc}^{-1}$ as the pivot. 
The Fisher matrix calculation for the errors in the parameters also
assumes a ``true'' model around which the derivatives are taken; for
this we use a ${\Lambda}$CDM model with $\Omega_b h^2 =
0.0222$, $\Omega_m h^2 = 0.136$, $\Omega_\Lambda = 0.73$, $\tau = 0.1$, and
a flat primordial power spectrum.

\subsection{Lyman-$\alpha$}

For the Lyman-$\alpha$ data, the error bars that were reported in
\cite{croft} for the linear matter power spectrum are used.  The
primordial power spectrum is related to the linear matter power
spectrum by 
\begin{equation}
	P_\mathrm{LM}(k) = P_0 k {A_S}^2(k) T^2(k) , 
\end{equation}
where $T^2(k)$ is the transfer function and contains the dependence upon cosmological parameters, and $P_0$ is a normalizing constant.  
Then the error bars for the power spectrum parameters are calculated via standard error propagation techniques using the previous equation and the analytic form for the transfer function \cite{pd}
\begin{eqnarray}
	T(q) &=& \frac{\ln(1+2.34\,q)}{2.34\, q} \nonumber \\
	&& \times \left[ 1 + aq + (bq)^2 + (cq)^3 + (dq)^4 	\right]^{-\frac14} , 
\end{eqnarray}
where $q \equiv k / \Omega h^2 \exp(-2\Omega_b)$ and $a,b,c,d$ are fit parameters which are irrelevant to the error analysis.  We use the results of the CMB parameter
estimation as inputs for determining the errors in $h$, $\Omega$, and $\Omega_b$.

The large systematic normalization error reported in \cite{croft} is a problem for estimating primordial power spectrum amplitudes, but does not affect local estimates of the slope or higher derivatives, so we do not include it.

\subsection{Error contours for current and future data
  \label{ssec:error_contours}} 

\begin{figure}
	\includegraphics[width=3.4in]{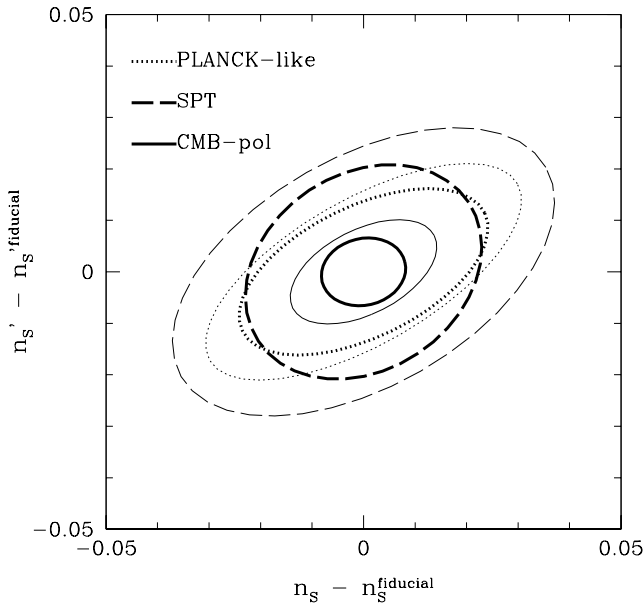}
	\caption{
		\label{fig:allovals}
		68\% confidence regions in the $n_S$--${n_S}'$ plane for various experiments
		of the sort described in table \ref{tab:exp}.  The SPT constraints are
		shown dashed, PLANCK-like dotted, and CMB-pol constraints solid.
		We use prior I-6 as described in section \ref{sec:results}.
		The thin lines are constraints without polarization information.
	}
\end{figure} 
\begin{figure}
	\includegraphics[width=3.4in]{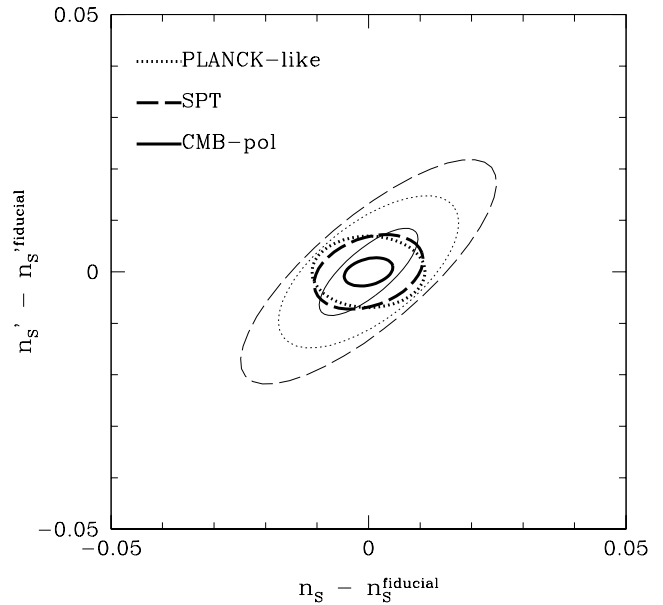}
	\caption{
		\label{fig:priorovals}
		68\% confidence regions in the $n_S$--${n_S}'$ plane for the same experiments
		as in figure \ref{fig:allovals}, only now using prior IIb as described in
		section \ref{sec:results}.
	}
\end{figure}
		
\begin{figure}
	\includegraphics[width=3.4in]{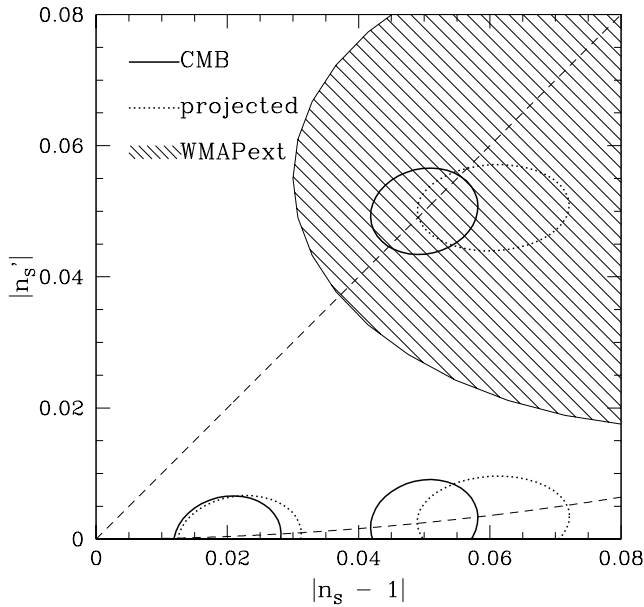}
	\caption{
		\label{fig:cmbovals}
		Solid contours show constraints in the $n_S$--${n_S}'$
		plane from the CMB data of 
		the ``CMB-pol'' experiment of table \ref{tab:exp} (using
		prior I-6) 
		for scales near $k = 0.0565$.
		Several different fiducial models are shown, two
		representative of the types 		shown in
		figure \ref{fig:models} and 
		one large-${n_S}'$ model of the type described in
		\cite{stew1}.
		The dotted contours are the result of assuming the
		consistency relation of 
		equation \ref{eq:dconsistency} and extrapolating from CMB
		scales to Lyman-$\alpha$ scales ($k = 2.39$).  If the 
		relation is valid, constraints from the two types of
		data will be correlated in this plane as illustrated
		by the pairs of adjacent solid and dashed contours.
		All contours are drawn enclosing regions of 68\% confidence level.
	}
\end{figure}
\begin{table} 
	\caption{\label{tab:exp} Noise parameters used for simulating various experiments.}
	\begin{ruledtabular}
	\begin{tabular}{ccccc}
	Experiment & $\sigma_T $ & $\beam$ & $\fsky$ & $\ell_\textrm{max}$ \\
	\hline
	WMAP-like & 20 $\mu$K& 18' & 0.7 & 1500 \\
	PLANCK-like & 10 $\mu$K& 6'& 0.7 & 3000 \\
	SPT & 12 $\mu$K & 0.9' & 0.1 & 4000 \\
	CMB-pol & 3 $\mu$K& 3' & 0.7 & 3000 \\

	\end{tabular}
	\end{ruledtabular}
\end{table}

An illustration of CMB and Lyman-$\alpha$ constraints on the
primordial power spectrum is shown in figures \ref{fig:current} and
\ref{fig:super}. 
If we fit a function to all the data points, and assume that function
to be linear then of course the slope will be tightly constrained.  If
we allow the function to have a more complicated shape, the slope at
any point becomes less well constrained.  Figure \ref{fig:current}
roughly represents current experimental limits.  The main point of
this paper is that future data can become good enough to loosen the assumptions on the
shape and still produce very tight constraints.  Figure 
\ref{fig:super} gives an illustration, by showing constraints on the
binned power spectrum form 
some future experiments accurate enough to clearly distinguish a model
with ${n_S}' = 0.05$ 
from a model with ${n_S}' = (0.05)^2$, even  using data spanning
only one order of magnitude in wavenumber. 

The ultimate limiting factor in how precise all these measurements can
be is due to cosmic variance.  For the CMB, the fractional error from
this effect is $\Delta C_\ell / C_\ell \sim 1 / \sqrt{\ell}$, which
means even for large $\ell$ each individual $C_\ell$ can only ever be
known to within a few percent.   
Thus the error in figure \ref{fig:super} is mostly cosmic variance
limited, and the only way to further reduce the error is to assume
some smoothness for the primordial power spectrum and bin the data.
Figures \ref{fig:current} and \ref{fig:super} use a step-wise
parametrization to constrain the primordial 
power spectrum in bins in $\ln k$, which is meaningful as long as the
power spectrum is a smooth function of wavenumber relative to the bin
width.  For the best CMB data points, which here occur near scales
corresponding to $\ell \sim 1000$, the bin width is roughly equivalent
to binning a few hundred $C_\ell$ together.  

From figure \ref{fig:super} we can see that the CMB accurately probes
the primordial power spectrum at somewhat larger scales than those
where the Lyman-$\alpha$ data is most accurate.  This means the two
experiments provide constraints on $n_S$ and ${n_S}'$ at different
scales, allowing us to further test our models by looking at how they
predict these quantities should change with wavenumber.  We will
explore this idea further in section \ref{sec:results}. 

To then make error contours in the $n_S$--${n_S}'$ plane, we use the
covariance matrix from the parameter analysis to marginalize over
other parameters and determine the covariance matrix for just $n_S$ and ${n_S}'$.
Of the four hypothetical CMB experiments listed in table \ref{tab:exp}, we show how
well the last three place constraints in the $n_S$--${n_S}'$ plane
(for two different priors) in figures \ref{fig:allovals} and
\ref{fig:priorovals}.  We examine the fourth experiment (the best) in
detail (using the weakest prior) in figure \ref{fig:cmbovals}.  These
results are discussed at length in the following section.  For
hypothetical  Lyman-$\alpha$ experiments, we do not understand the
physics connecting 
the primordial power spectrum to measurements as well as for the CMB. 
We therefore ``simulate'' improvements in experiments as an overall reduction in
statistical uncertainty due to larger samples, and an improved knowledge of the
transfer function due to decreased errors in $h$, $\Omega$, and $\Omega_b$.
For Lyman-$\alpha$ experiments to provide constraints competitive with those expected from Planck will require datasets roughly 100 times larger than current ones, and more importantly, an understanding of systematics (or at least those that affect estimates of $n_S$ and ${n_S}'$) down to the percent level.  Given that these systematics represent a lack of understanding of the (scale-dependent) light-to-mass and baryon-to-dark matter ratios, such an improvement may not appear soon.
We hope to study the error for future Lyman-$\alpha$ experiments in more
realistic detail in future work.

\section{Testing Inflation}
\label{sec:results}
\subsection{Results}

\begin{table}
	\caption{\label{tab:prior1}Marginalized errors ($\times 1000$) in $n_S$ and
	${n_S}'$ for
	simulated experiments with different priors on truncating the Taylor expansion of the
	spectral index.  Power spectrum parameters are amplitude, tilt ($n_S$), and
	for column I-$m$, the first $m$ derivatives of the tilt.}
	\begin{tabular}{|r||r@{, }l|r@{, }l|r@{, }l|}
	\hline
	& \multicolumn{2}{c|}{prior I-6} &
	\multicolumn{2}{c|}{prior I-3} &
	\multicolumn{2}{c|}{prior I-1} \\
	\multicolumn{1}{|c||}{Experiment} & $\delta n_S$ & $\delta{n_S}'$ 
	& $\delta n_S$ & $\delta{n_S}'$
	& $\delta n_S$ & $\delta{n_S}'$ \\
	& \multicolumn{2}{c|}{$(\times 1000$)} &
	  \multicolumn{2}{c|}{$(\times 1000$)} &
	  \multicolumn{2}{c|}{$(\times 1000$)} \\
	\hline\hline
	WMAP-like & 309 & 274 & 258 & 152 & 69.1 & 33.9\\
	w polarization & 203 & 260 & 195 & 132 & 67.8 & 20.6\\
	\hline
	PLANCK-like & 20.3 & 13.9 & 16.6 & 10.7 & 11.6 & 9.79 \\
	w polarization & 16.0 & 10.7 & 12.4 & 7.48 & 6.90 & 3.94 \\
	\hline
	SPT & 24.6 & 18.6 & 21.5 & 14.6 & 16.3 & 14.5 \\
	w polarization & 15.2 & 13.8 & 10.7 & 8.67 & 6.37 & 4.78 \\
	\hline
	CMB-pol & 9.43 & 6.71 & 8.28 & 5.67 & 6.17 & 5.60 \\
	w polarization & 5.42 & 4.37 & 4.08 & 2.42 & 2.32 & 1.75 \\
	\hline
	\end{tabular}
\end{table}

\begin{table}
	\caption{\label{tab:prior2}Marginalized errors ($\times 1000$) in $n_S$ and
	${n_S}'$ for
	simulated experiments with different priors on the higher derivatives of
	the spectral index.  Prior I-6 from Table \ref{tab:prior1} above is for reference.
	Priors IIa and IIb use all six derivatives in the joint parameter estimation, but
	IIa imposes the constraint that all higher (${n_S}''$ and above) derivatives are
	within $0.01$ of scale invariant, and IIb imposes the
	consistency relation (Eqn. \ref{eq:dconsistency}) on the higher derivatives.
	}
	\begin{tabular}{|r||r@{, }l|r@{, }l|r@{, }l|}
	\hline
	& \multicolumn{2}{c|}{prior I-6} &
	\multicolumn{2}{c|}{prior IIa} &
	\multicolumn{2}{c|}{prior IIb} \\
	\multicolumn{1}{|c||}{Experiment} & $\delta n_S$ & $\delta{n_S}'$
	& $\delta n_S$ & $\delta{n_S}'$
	& $\delta n_S$ & $\delta{n_S}'$ \\
	& \multicolumn{2}{c|}{$(\times 1000$)} &
		\multicolumn{2}{c|}{$(\times 1000$)} &
		\multicolumn{2}{c|}{$(\times 1000$)} \\
	\hline\hline
	WMAP-like & 309 & 274 & 97.8 & 38.9 & 70.6 & 34.6 \\
	w polarization & 203 & 260 & 93.1 & 34.1 & 69.4 & 22.4 \\
	\hline
	PLANCK-like & 20.3 & 13.9 & 12.3 & 10.1 & 11.6 & 9.79 \\
	w polarization & 16.0 & 10.7 & 7.50 & 5.14 & 7.26 & 4.60 \\
	\hline
	SPT & 24.6 & 18.6 & 16.6 & 14.5 & 16.5 & 14.5 \\
	w polarization & 15.2 & 13.8 & 7.36 & 5.19 & 7.00 & 4.80 \\
	\hline
	CMB-pol & 9.43 & 6.71 & 6.81 & 5.72 & 6.36 & 5.61 \\
	w polarization & 5.42 & 4.37 & 3.38 & 2.26 & 3.13 & 1.82 \\
	\hline
	\end{tabular}
\end{table}
\begin{figure}
	\includegraphics[width=3.4in]{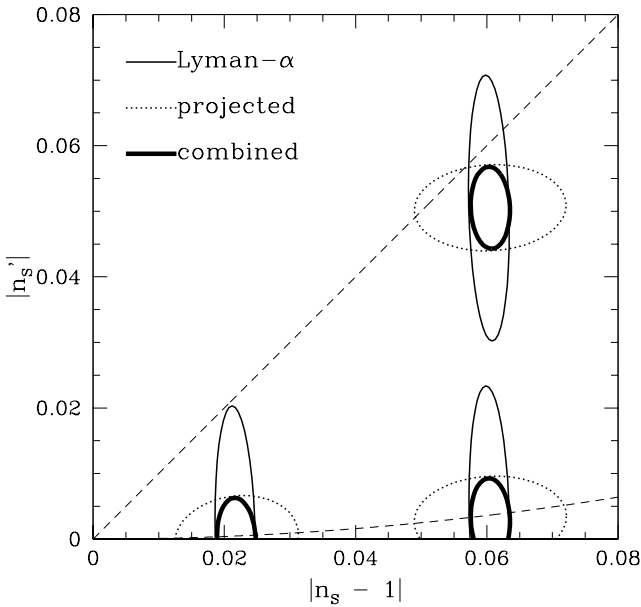}
	\caption{
		\label{fig:lyaovals}
		Constraints in the $n_S$--${n_S}'$ plane from the Lyman-$\alpha$ data of
		figure \ref{fig:super} for scales near $k = 2.39$ (thin large ovals).  The
		projected ovals from CMB scales (dotted ovals, from Fig. \ref{fig:cmbovals})
		are then combined
		with the Lyman-$\alpha$ data to form improved constraints (thick small ovals).
		Again, 68\% confidence level regions are plotted.
	}
\end{figure}

For completeness, we discuss a range of possible priors, each of which
represents a different points of view on what one wants to take
as an assumption and what one is trying to test.  In tables
\ref{tab:prior1} and \ref{tab:prior2} we report the constraints that 
various CMB experiments can place on $n_S$ and ${n_S}'$.  Our weakest
prior, which we refer to as I-6, is 
simply to use the four cosmological parameters $(\ell_A, \Omega_b h^2, \Omega_m h^2,
\tau)$ and eight power spectrum parameters ($P$ and the first
seven coefficients in the expansion of $n_S$).  Having so many parameters for
the power spectrum allows the shape to vary quite a bit, and loosens constraints
on each term of the expansion.
We include so many parameters not so much because constraints on all of them will be interesting (some, in fact, will probably always be unmeasurable), but to show the effect various assumptions about them will have on the constraints on $n_S$ and ${n_S}'$.  If the reader dislikes these parameters, our prior I-1 is equivalent to not including them at all.

To get better constraints requires either using a more restrictive prior or
improving the experiment.  In table \ref{tab:prior1} we change the prior by using
fewer power spectrum parameters.  Prior I-3 uses only the first four terms of the
expansion of $n_S$ (i.e. up to third order), and prior I-1 uses only the first two
terms, such that the only power spectrum parameters for prior I-1 are amplitude,
$n_S$, and ${n_S}'$.  In table \ref{tab:prior2} we change the prior by placing
a priori constraints on the higher derivative terms of the expansion, rather than
dropping them completely.  Prior IIa supposes that all the higher derivative terms
(${n_S}''$ and higher) are ``small'', less than $0.01$.  Prior IIb also imposes
the constraint that the higher derivatives be ``small'', but supposes they
fall off in the form of the consistency relation of equation
\ref{eq:dconsistency}.  

Prior I-6 represents the weakest assumptions, and prior I-1 the
tightest assumptions. (Prior I-1 basically adds just one new parameter,
${n_S}'$, to the canonical set.)  Prior IIb represents an assumption
of the standard inflationary picture for the higher derivatives, but
places no prior constraint in the $n_S-{n_S}'$ plane so those
parameters can be used to test the standard picture. (Of course, if the
the standard slow roll picture fails, prior IIb may no longer be of interest)

In the time since the preprint of this paper first appeared, the WMAP
collaboration announced their results \cite{wmap1}.  
Our predicted error for $n_S$ and ${n_S}'$ of 0.069 and 0.034 matches up quite
well to their reported errors of 0.060 and 0.038,  (using our prior I-1,
which most closely 
matches the ``WMAPext'' analysis of \cite{wmap2}.)\footnote{Our
characterization of the WMAP noise and beam size is somewhat more
pessimistic than their reported numbers, but we suspect this is
compensated by further experimental details we do not include.} While the
WMAP results are not statistically
very significant for our purposes, we 
plot the error contour in figure \ref{fig:cmbovals} for comparison.
(Note that if the central value does not change much as the data
improve the implications for inflation will be very interesting.)
 
We have simulated CMB experiments with and without polarization measurements.
Since the primordial power spectrum affects the CMB temperature and
polarization in exactly the same way, naively polarization simply adds
a second way of measuring the same thing and should only reduce
uncertainty by a factor of $\sqrt 2$.   
However, allowing joint estimation of other cosmological parameters in
addition to those describing the shape of the primordial power
spectrum introduces confusion and near-degeneracies.  The value of
measuring polarization is not so much that it directly puts limits on
the power spectrum, but in reducing the confusion with other
parameters.  Our analysis shows the increasing value of polarization as CMB experiments improve.  


In figures \ref{fig:cmbovals} and \ref{fig:lyaovals} we see the error
contours in the $n_S$--${n_S}'$  plane for a hypothetical super-Planck
CMB temperature and polarization experiment and for a ``super''
Ly-$\alpha$ survey.  We see that such data would provide very
significant constraints in this space. In particular, these
experiments are good enough to clearly distinguish points on the line
${n_S}' = n_S-1$ from the line ${n_S}' = (n_S-1)^2$ for all but very
small values of $n_S$, and thus would offer significant tests of the
standard inflationary picture.

Combining data from both experiments will provide additional
constraints and tests.  Each experiment provides constraints 
in the $n_S-{n_S}'$ plane, but on somewhat different scales.  These
amount to providing constraints on the inflaton 
potential $V(\phi)$ near a particular wavenumber $k$.  There are several
possible options for combining data from several experiments.  One approach
is to use a single parametrization for the primordial power spectrum
and then to jointly estimate all parameters using the full dataset.
If both experiments were at the same scale, this would amount to
simply overlapping their individual error contours.  
To perform joint estimation for experiments at different scales, we would
want to find parameters that are ``good'' across different scales, but this conflicts
with our aim to test how good (i.e. constant) a parameter $n_S$ really is across
a large range of scales.
Also, a more immediate concern, is that
different experiments often have a (sometimes poorly characterized) systematic
error in their relative normalization which causes problems for a joint
parameter analysis.  For a recent discussion of these issues, see \cite{I03}.

For simplicity, then, we choose a second option, which is to look at the 
two experiments separately and then use the 
consistency
equation (Eqn. \ref{eq:dconsistency}) to produce the inequality 
\begin{equation}
	\frac{n_S(k_\mathrm{Ly\alpha}) - n_S(k_\mathrm{CMB}) }
		{\ln k_\mathrm{Ly\alpha} - \ln k_\mathrm{CMB} }
	<
	(n_S(k_\mathrm{Ly\alpha}) - 1)^2  
\end{equation}
as an additional inflationary test.  By looking at only derivatives of the 
power spectrum we avoid the relative normalization problems.
Visually, this amounts to projecting the CMB contours for $n_S$ and
${n_S}'$ up to Lyman-$\alpha$ scales (or vice-versa)
and checking to see if the contours overlap, as shown in figure \ref{fig:lyaovals}.
This third test is not redundant because we use different pivot points
($k_\mathrm{Ly\alpha}$ and $k_\mathrm{CMB}$) for the different
datasets.  A potential with large higher derivatives could pass the ${n_S}' < (n_S
-1)^2$ test at a particular scale and yet fail it when data from
different scales is used.

Finally, we would like to remind the reader that for figures
\ref{fig:cmbovals} and \ref{fig:lyaovals}, the real information is in
the size of the error contours rather than their actual placement.  
In the absence of real data, we show contours from simulated data for a
small sample of models all of which show correlations between the two
experiments consistent with the standard inflationary picture.
Ultimately nature will tell us if such correlations are really there.  

\subsection{Projected errors and cross-correlations}
\label{ssec:confusion}
\begin{figure}
	\includegraphics[width=3.4in]{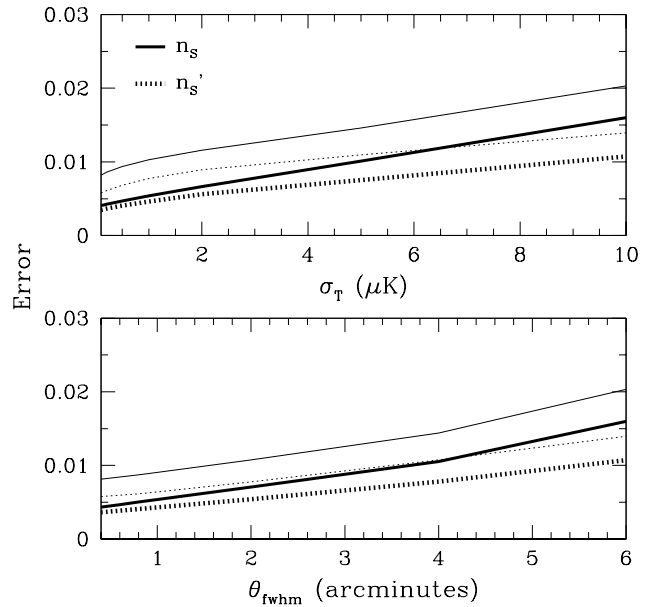}
	\caption{
		\label{fig:noise}
		Marginalized (one-sigma) errors of $n_S$ (solid line)
		and ${n_S}'$ (dashed  
		line) as a function of pixel noise and beam width.
		Values with(without) 
		polarization are shown as thick(thin) lines.  For the
		lower plot, beam 
		width has been fixed at $6'$, and for the upper
		noise has been fixed at 
		$10 \mu$K.
	}
\end{figure}

To further show the value of better experiments, we have investigated how the
projected errors in $n_S$ and ${n_S}'$ should change as a result of improving
both resolution and noise levels.  Figure \ref{fig:noise} shows the
marginalized errors (with prior I-6) with and without polarization, as functions
of beam width and pixel noise.

A source of worry is possible cross-correlation between our power spectrum parameters
and other cosmological parameters.  Confusion between tilt and optical depth is
well-known, and polarization helps greatly at reducing such confusion.
We did find that there is generally a correlation
between $\ell_A$ and our power spectrum parameters (of which ${n_S}'$ is of
the most interest).  The correlation seems to arise because power spectrum parameters can combine to mimic a slight horizontal shift of a peak.  Measuring the locations of multiple peaks makes this conspiracy of power spectrum parameters more difficult, however, and for the best experiments the correlation disappears.  For the near future, an accurate determination of the angular scale of the sound horizon at last scattering will be important for placing constraints on inflationary parameters.

\subsection{The Fisher matrix approximation}
\label{ssec:fisher}

One possible source of error in our calculations stems from the
approximations that go into 
the Fisher matrix analysis technique.  For low $\ell$, the covariance
matrix $\mathbf C$ is dominated by cosmic variance, and equation
\ref{eq:fisher} can be rewritten in terms of a new variable $Z_\ell
\equiv \ln C_\ell$.  As pointed out by Bond et.~al.~\cite{bjk}, 
these $Z_\ell$ are better variables than the $C_\ell$ for Fisher
analysis in the 
sense that they are Gaussian distributed.  For large enough $\ell$, the difference between using $Z_\ell$ and $C_\ell$ is minor, however.

Just having variables that are Gaussian distributed is not enough; the
variables should respond linearly to the parameters.  Linearity in
response to the cosmological parameters is a well-studied problem and
has been discussed extensively in the literature (see \cite{kmj} for a
recent discussion).  What remains is to check our power spectrum
parameters.  We have done this by examining $Z_\ell$ as a function of
the power spectrum parameters. 

For the power spectrum parameters we are most interested in ($n_S$ and
${n_S}'$), deviations from linearity are less than one percent for
parameter values within all the errorbars that we report in tables
\ref{tab:prior1} and \ref{tab:prior2}.  For the higher order
parameters, the worst deviations from linearity occur only for small
$\ell$ ($\sim 10$), for which we found the linear approximation to be
valid within one percent for parameter values less than around 0.02.
In our calculations, we varied the higher order parameters by amounts
several times smaller than this, and while the resulting errorbars
were sometimes larger than this, at these small $\ell$ the error from
cosmic variance is also large, so the effect on the overall parameter
analysis is small.  We checked this by redoing the parameter analysis
without including low $\ell$ values at all, and the qualitative
results of our work do not change.   For higher $\ell$ ($\sim 200$ and
above), the linear approximation remains good up to parameter values
of order unity. 

\section{Conclusion}
\label{sec:conclusion}

We have shown that the next generation of cosmological experiments
should determine the shape of the primordial power
spectrum sufficiently to allow new tests of the the
standard picture of inflation.  If the standard picture is upheld, a
new level of differentiation among different inflaton potentials will
be possible.  We have investigated the potential impact of new data on
both the CMB and the Lyman-$\alpha$ forest.

The Lyman-$\alpha$ data offers a promising route to testing slow-roll
models both on its own, and in conjunction with CMB data.  Currently
published data does not get too far with this enterprise, but
next-generation observation could have considerable impact\footnote{As this
work was completed we learned that the Sloan Digital Sky Survey is
preparing to release a new Lyman-$\alpha$ dataset.  While not as large 
as our survey which is ``next generation''Lyman-$\alpha$ dataset
  simulated in this paper, it may have considerable importance to the
  issues raised in this article \cite{uros}}.  

For the CMB data, if we are interested in general constraints without
placing restrictive 
priors on the primordial power spectrum,
Planck-like CMB experiments do not quite have the precision
to put the
strong limits on ${n_S}'$ that we desire.  It is not until relatively
high $\ell$ that the uncertainty from cosmological variance is low
enough for our requirements, and it is precisely at these high $\ell$
that the Planck experiment noise rapidly becomes dominant so a
further improvement beyond
Planck is needed. Our hypothetical ``CMB-pol'' experiment should start
placing interesting constraints in the $n_S$--${n_S}'$ plane. 
Also important is the measurement of the (E-mode)
polarization channel, which is vital to reducing degeneracies that
make the tests more challenging.

For constraints on the power
spectrum itself (as opposed to other cosmological parameters), information from
low multipole moments ($\ell < 500$) contributes very little due to
cosmic variance.  However, coverage of a reasonable fraction of the sky is needed to retain high resolution in $\ell$, and simply to beat down statistical noise.  The proposed South Pole Telescope (SPT) may do well in this regard.
Higher multipole moments are useful up until Silk damping reduces
the overall CMB signal.
As CMB experiments improve, polarization will become more important
as the key to breaking degeneracies between the effects of the power
spectrum shape, which affects temperature and polarization identically, and other cosmological parameters, which generally do not.

\section*{Acknowledgments}
We thank Lloyd Knox, Manoj Kaplinghat, and David Spergel for helpful conversations.
We were supported in part by DOE grant DE-FG03-91ER40674. 

\appendix*
\section{Models}
\label{appendixA}
The potentials used in the models shown in figure \ref{fig:models}, from left to right are (in units where $M_P \equiv 1$)
\begin{eqnarray}
	M^4 \left[1 + A \ln \phi \right] , \\
	M^4 \left[1 - e^{-\phi/2} \right]^2 , \\
	M^4 \left[1 - e^{-\phi} \right] , \\
	M^4 \left[1 + A \phi^2 \right] , \\
	M^4 \cos^2 \frac{\phi}A  ,\\
	M^4 \left[1 - A \phi^{12} \right] , \\
	M^4 \left[1 - A \phi^2 \right] , \\
	M^4 \left[1 - A \phi^4 \right] , \\
	M^4 \left[1 + \frac12 \phi^2 \left( A + 0.6 \right) \right] .
\end{eqnarray}
In all cases the mass scale and other parameters were chosen as in
table \ref{tab:mod} to produce a $\delta\rho /  \rho$ of roughly $2
\times 10^{-5}$. 
\begin{table}[h] 
	\caption{\label{tab:mod} Inflationary model parameters}
	\begin{ruledtabular}
	\begin{tabular}{ccc}
	Model \# & $M$ & $A$ \\
	\hline
	1 & $10^{-3}$ & $5 \times 10^{-2}$ \\
	2 & $10^{-3}$ & --- \\
	3 & $10^{-3}$ & --- \\
	4 & $10^{-4}$ & $10^{-3}$ \\
	5 & $10^{-9/4}$ & $100$ \\
	6 & $2 \times 10^{-4}$ & $5 \times 10^{-3}$ \\
	7 & $10^{-3}$ & $10^{-2}$ \\
	8 & $10^{-6}$ & $\frac14 \times 10^{9}$ \\
	9 & $10^{-16}$ & $0.139$

	\end{tabular}
	\end{ruledtabular}
\end{table}

\bibliography{nextgen.bib}

\begin{thebibliography}{36}
\expandafter\ifx\csname natexlab\endcsname\relax\def\natexlab#1{#1}\fi
\expandafter\ifx\csname bibnamefont\endcsname\relax
  \def\bibnamefont#1{#1}\fi
\expandafter\ifx\csname bibfnamefont\endcsname\relax
  \def\bibfnamefont#1{#1}\fi
\expandafter\ifx\csname citenamefont\endcsname\relax
  \def\citenamefont#1{#1}\fi
\expandafter\ifx\csname url\endcsname\relax
  \def\url#1{\texttt{#1}}\fi
\expandafter\ifx\csname urlprefix\endcsname\relax\def\urlprefix{URL }\fi
\providecommand{\bibinfo}[2]{#2}
\providecommand{\eprint}[2][]{\url{#2}}

\bibitem[{\citenamefont{Copeland et~al.}(1993)\citenamefont{Copeland, Kolb,
  Liddle, and Lidsey}}]{ckll}
\bibinfo{author}{\bibfnamefont{E.}~\bibnamefont{Copeland}},
  \bibinfo{author}{\bibfnamefont{E.~W.} \bibnamefont{Kolb}},
  \bibinfo{author}{\bibfnamefont{A.~R.} \bibnamefont{Liddle}},
  \bibnamefont{and} \bibinfo{author}{\bibfnamefont{J.~E.}
  \bibnamefont{Lidsey}}, \bibinfo{journal}{Phys.\ Rev.\ D}
  \textbf{\bibinfo{volume}{48}}, \bibinfo{pages}{2529} (\bibinfo{year}{1993}).

\bibitem[{\citenamefont{Knox}(1995)}]{lk}
\bibinfo{author}{\bibfnamefont{L.}~\bibnamefont{Knox}},
  \bibinfo{journal}{Phys.\ Rev.\ D} \textbf{\bibinfo{volume}{52}},
  \bibinfo{pages}{4307} (\bibinfo{year}{1995}).

\bibitem[{\citenamefont{Kosowsky and Turner}(1995)}]{kt}
\bibinfo{author}{\bibfnamefont{A.}~\bibnamefont{Kosowsky}} \bibnamefont{and}
  \bibinfo{author}{\bibfnamefont{M.~S.} \bibnamefont{Turner}},
  \bibinfo{journal}{Phys.\ Rev.\ D} \textbf{\bibinfo{volume}{52}},
  \bibinfo{pages}{1739} (\bibinfo{year}{1995}).

\bibitem[{\citenamefont{Jungman et~al.}(1996)\citenamefont{Jungman,
  Kamionkowski, Kosowsky, and Spergel}}]{jkks}
\bibinfo{author}{\bibfnamefont{G.}~\bibnamefont{Jungman}},
  \bibinfo{author}{\bibfnamefont{M.}~\bibnamefont{Kamionkowski}},
  \bibinfo{author}{\bibfnamefont{A.}~\bibnamefont{Kosowsky}}, \bibnamefont{and}
  \bibinfo{author}{\bibfnamefont{D.~N.} \bibnamefont{Spergel}},
  \bibinfo{journal}{Phys.\ Rev.\ D} \textbf{\bibinfo{volume}{54}},
  \bibinfo{pages}{1332} (\bibinfo{year}{1996}).

\bibitem[{\citenamefont{{Copeland} et~al.}(1998)\citenamefont{{Copeland},
  {Grivell}, and {Liddle}}}]{cgl}
\bibinfo{author}{\bibfnamefont{E.~J.} \bibnamefont{{Copeland}}},
  \bibinfo{author}{\bibfnamefont{I.~J.} \bibnamefont{{Grivell}}},
  \bibnamefont{and} \bibinfo{author}{\bibfnamefont{A.~R.}
  \bibnamefont{{Liddle}}}, \bibinfo{journal}{Mon. Not. Roy. Astron. Soc.}
  \textbf{\bibinfo{volume}{298}}, \bibinfo{pages}{1233} (\bibinfo{year}{1998}).

\bibitem[{\citenamefont{Dodelson et~al.}(1997)\citenamefont{Dodelson, Kinney,
  and Kolb}}]{dkk}
\bibinfo{author}{\bibfnamefont{S.}~\bibnamefont{Dodelson}},
  \bibinfo{author}{\bibfnamefont{W.~H.} \bibnamefont{Kinney}},
  \bibnamefont{and} \bibinfo{author}{\bibfnamefont{E.~W.} \bibnamefont{Kolb}},
  \bibinfo{journal}{Phys.\ Rev.\ D} \textbf{\bibinfo{volume}{56}},
  \bibinfo{pages}{3207} (\bibinfo{year}{1997}).

\bibitem[{\citenamefont{Kinney et~al.}(2001)\citenamefont{Kinney, Melchiorri,
  and Riotto}}]{kmr}
\bibinfo{author}{\bibfnamefont{W.~H.} \bibnamefont{Kinney}},
  \bibinfo{author}{\bibfnamefont{A.}~\bibnamefont{Melchiorri}},
  \bibnamefont{and} \bibinfo{author}{\bibfnamefont{A.}~\bibnamefont{Riotto}},
  \bibinfo{journal}{Phys.\ Rev.\ D} \textbf{\bibinfo{volume}{63}},
  \bibinfo{pages}{023505} (\bibinfo{year}{2001}).

\bibitem[{\citenamefont{{Hannestad} et~al.}(2002)\citenamefont{{Hannestad},
  {Hansen}, {Villante}, and {Hamilton}}}]{hann1}
\bibinfo{author}{\bibfnamefont{S.}~\bibnamefont{{Hannestad}}},
  \bibinfo{author}{\bibfnamefont{S.~H.} \bibnamefont{{Hansen}}},
  \bibinfo{author}{\bibfnamefont{F.~L.} \bibnamefont{{Villante}}},
  \bibnamefont{and} \bibinfo{author}{\bibfnamefont{A.~J.~S.}
  \bibnamefont{{Hamilton}}}, \bibinfo{journal}{Astroparticle Physics}
  \textbf{\bibinfo{volume}{17}}, \bibinfo{pages}{375} (\bibinfo{year}{2002}).

\bibitem[{\citenamefont{Hansen and Kunz}(2002)}]{hk}
\bibinfo{author}{\bibfnamefont{S.~H.} \bibnamefont{Hansen}} \bibnamefont{and}
  \bibinfo{author}{\bibfnamefont{M.}~\bibnamefont{Kunz}},
  \bibinfo{journal}{Mon. Not. Roy. Astron. Soc.}
  \textbf{\bibinfo{volume}{336}}, \bibinfo{pages}{1007} (\bibinfo{year}{2002}),
  \eprint{hep-ph/0109252}.

\bibitem[{\citenamefont{Bucher et~al.}(2000)\citenamefont{Bucher, Moodley, and
  Turok}}]{Bucher:2000cd}
\bibinfo{author}{\bibfnamefont{M.}~\bibnamefont{Bucher}},
  \bibinfo{author}{\bibfnamefont{K.}~\bibnamefont{Moodley}}, \bibnamefont{and}
  \bibinfo{author}{\bibfnamefont{N.}~\bibnamefont{Turok}}
  (\bibinfo{year}{2000}), \eprint{astro-ph/0011025}.

\bibitem[{\citenamefont{Guth}(1981)}]{guth}
\bibinfo{author}{\bibfnamefont{A.~H.} \bibnamefont{Guth}},
  \bibinfo{journal}{Phys.\ Rev.\ D} \textbf{\bibinfo{volume}{23}},
  \bibinfo{pages}{347} (\bibinfo{year}{1981}).

\bibitem[{\citenamefont{Albrecht and Steinhardt}(1982)}]{as}
\bibinfo{author}{\bibfnamefont{A.}~\bibnamefont{Albrecht}} \bibnamefont{and}
  \bibinfo{author}{\bibfnamefont{P.}~\bibnamefont{Steinhardt}},
  \bibinfo{journal}{Phys.\ Rev.\ Lett.} \textbf{\bibinfo{volume}{48}},
  \bibinfo{pages}{1220} (\bibinfo{year}{1982}).

\bibitem[{\citenamefont{Linde}(1982)}]{linde}
\bibinfo{author}{\bibfnamefont{A.~D.} \bibnamefont{Linde}},
  \bibinfo{journal}{Phys.\ Lett.\ B} \textbf{\bibinfo{volume}{108}},
  \bibinfo{pages}{389} (\bibinfo{year}{1982}).

\bibitem[{\citenamefont{Linde}(1983)}]{lindechaos}
\bibinfo{author}{\bibfnamefont{A.~D.} \bibnamefont{Linde}},
  \bibinfo{journal}{Phys.\ Lett.\ B} \textbf{\bibinfo{volume}{129}},
  \bibinfo{pages}{177} (\bibinfo{year}{1983}).

\bibitem[{\citenamefont{Bardeen}(1980)}]{bard}
\bibinfo{author}{\bibfnamefont{J.~M.} \bibnamefont{Bardeen}},
  \bibinfo{journal}{Phys.\ Rev.\ D} \textbf{\bibinfo{volume}{22}},
  \bibinfo{pages}{1882} (\bibinfo{year}{1980}).

\bibitem[{\citenamefont{Dodelson and Stewart}(2002)}]{stew1}
\bibinfo{author}{\bibfnamefont{S.}~\bibnamefont{Dodelson}} \bibnamefont{and}
  \bibinfo{author}{\bibfnamefont{E.}~\bibnamefont{Stewart}},
  \bibinfo{journal}{Phys. Rev.} \textbf{\bibinfo{volume}{D65}},
  \bibinfo{pages}{101301} (\bibinfo{year}{2002}), \eprint{astro-ph/0109354}.

\bibitem[{\citenamefont{Tegmark and Zaldarriaga}(2002)}]{tz}
\bibinfo{author}{\bibfnamefont{M.}~\bibnamefont{Tegmark}} \bibnamefont{and}
  \bibinfo{author}{\bibfnamefont{M.}~\bibnamefont{Zaldarriaga}},
  \bibinfo{journal}{Phys.\ Rev.\ D} \textbf{\bibinfo{volume}{66}},
  \bibinfo{pages}{103508} (\bibinfo{year}{2002}).

\bibitem[{\citenamefont{Lyth}()}]{lyth}
\bibinfo{author}{\bibfnamefont{D.~H.} \bibnamefont{Lyth}},
  \eprint{hep-ph/9609431}.

\bibitem[{\citenamefont{Liddle and Lyth}(2000)}]{ll}
\bibinfo{author}{\bibfnamefont{A.~R.} \bibnamefont{Liddle}} \bibnamefont{and}
  \bibinfo{author}{\bibfnamefont{D.~H.} \bibnamefont{Lyth}},
  \emph{\bibinfo{title}{Cosmological Inflation and Large-Scale Structure}}
  (\bibinfo{publisher}{Cambridge University Press}, \bibinfo{year}{2000}).

\bibitem[{\citenamefont{Dvali and Tye}(1999)}]{dt}
\bibinfo{author}{\bibfnamefont{G.}~\bibnamefont{Dvali}} \bibnamefont{and}
  \bibinfo{author}{\bibfnamefont{S.-H.~H.} \bibnamefont{Tye}},
  \bibinfo{journal}{Phys.\ Lett.\ B} \textbf{\bibinfo{volume}{450}},
  \bibinfo{pages}{72} (\bibinfo{year}{1999}).

\bibitem[{\citenamefont{Guth and Pi}(1982)}]{guthpi}
\bibinfo{author}{\bibfnamefont{A.~H.} \bibnamefont{Guth}} \bibnamefont{and}
  \bibinfo{author}{\bibfnamefont{S.-Y.} \bibnamefont{Pi}},
  \bibinfo{journal}{Physical Review Letters} \textbf{\bibinfo{volume}{49}},
  \bibinfo{pages}{1110} (\bibinfo{year}{1982}).

\bibitem[{\citenamefont{Shiu and Tye}(2001)}]{brane1}
\bibinfo{author}{\bibfnamefont{G.}~\bibnamefont{Shiu}} \bibnamefont{and}
  \bibinfo{author}{\bibfnamefont{S.-H.~H.} \bibnamefont{Tye}},
  \bibinfo{journal}{Phys.\ Lett.\ B} \textbf{\bibinfo{volume}{516}},
  \bibinfo{pages}{421} (\bibinfo{year}{2001}).

\bibitem[{\citenamefont{{Covi} et~al.}(2003)\citenamefont{{Covi}, {Lyth}, and
  {Melchiorri}}}]{clm}
\bibinfo{author}{\bibfnamefont{L.}~\bibnamefont{{Covi}}},
  \bibinfo{author}{\bibfnamefont{D.~H.} \bibnamefont{{Lyth}}},
  \bibnamefont{and}
  \bibinfo{author}{\bibfnamefont{A.}~\bibnamefont{{Melchiorri}}},
  \bibinfo{journal}{\prd} \textbf{\bibinfo{volume}{67}}, \bibinfo{pages}{43507}
  (\bibinfo{year}{2003}).

\bibitem[{\citenamefont{{Croft} et~al.}(2002)\citenamefont{{Croft}, {Weinberg},
  {Bolte}, {Burles}, {Hernquist}, {Katz}, {Kirkman}, and {Tytler}}}]{croft}
\bibinfo{author}{\bibfnamefont{R.~A.~C.} \bibnamefont{{Croft}}},
  \bibinfo{author}{\bibfnamefont{D.~H.} \bibnamefont{{Weinberg}}},
  \bibinfo{author}{\bibfnamefont{M.}~\bibnamefont{{Bolte}}},
  \bibinfo{author}{\bibfnamefont{S.}~\bibnamefont{{Burles}}},
  \bibinfo{author}{\bibfnamefont{L.}~\bibnamefont{{Hernquist}}},
  \bibinfo{author}{\bibfnamefont{N.}~\bibnamefont{{Katz}}},
  \bibinfo{author}{\bibfnamefont{D.}~\bibnamefont{{Kirkman}}},
  \bibnamefont{and} \bibinfo{author}{\bibfnamefont{D.}~\bibnamefont{{Tytler}}},
  \bibinfo{journal}{\apj} \textbf{\bibinfo{volume}{581}}, \bibinfo{pages}{20}
  (\bibinfo{year}{2002}).

\bibitem[{\citenamefont{Wang et~al.}(1999)\citenamefont{Wang, Spergel, and
  Strauss}}]{wss}
\bibinfo{author}{\bibfnamefont{Y.}~\bibnamefont{Wang}},
  \bibinfo{author}{\bibfnamefont{D.~N.} \bibnamefont{Spergel}},
  \bibnamefont{and} \bibinfo{author}{\bibfnamefont{M.~A.}
  \bibnamefont{Strauss}}, \bibinfo{journal}{Astrophys.\ J.}
  \textbf{\bibinfo{volume}{20}}, \bibinfo{pages}{510} (\bibinfo{year}{1999}).

\bibitem[{\citenamefont{Zaldarriaga and Seljak}(1996)}]{CMBFAST}
\bibinfo{author}{\bibfnamefont{M.}~\bibnamefont{Zaldarriaga}} \bibnamefont{and}
  \bibinfo{author}{\bibfnamefont{U.}~\bibnamefont{Seljak}},
  \bibinfo{journal}{Astrophys.\ J.} \textbf{\bibinfo{volume}{469}},
  \bibinfo{pages}{437} (\bibinfo{year}{1996}).

\bibitem[{\citenamefont{Zaldarriaga and Seljak}(1997)}]{sz_pol}
\bibinfo{author}{\bibfnamefont{M.}~\bibnamefont{Zaldarriaga}} \bibnamefont{and}
  \bibinfo{author}{\bibfnamefont{U.}~\bibnamefont{Seljak}},
  \bibinfo{journal}{Phys. Rev.} \textbf{\bibinfo{volume}{D55}},
  \bibinfo{pages}{1830} (\bibinfo{year}{1997}), \eprint{astro-ph/9609170}.

\bibitem[{\citenamefont{Knox et~al.}(2002)\citenamefont{Knox, Skordis, and
  Kaplinghat}}]{dash}
\bibinfo{author}{\bibfnamefont{L.}~\bibnamefont{Knox}},
  \bibinfo{author}{\bibfnamefont{C.}~\bibnamefont{Skordis}}, \bibnamefont{and}
  \bibinfo{author}{\bibfnamefont{M.}~\bibnamefont{Kaplinghat}},
  \bibinfo{journal}{Astrophys.\ J.} \textbf{\bibinfo{volume}{578}},
  \bibinfo{pages}{665} (\bibinfo{year}{2002}), \eprint{astro-ph/0203413}.

\bibitem[{\citenamefont{Peacock and Dodds}(1994)}]{pd}
\bibinfo{author}{\bibfnamefont{J.~A.} \bibnamefont{Peacock}} \bibnamefont{and}
  \bibinfo{author}{\bibfnamefont{S.~J.} \bibnamefont{Dodds}},
  \bibinfo{journal}{Mon.\ Not.\ R.\ Astron.\ Soc.}
  \textbf{\bibinfo{volume}{267}}, \bibinfo{pages}{1020} (\bibinfo{year}{1994}).

\bibitem[{\citenamefont{Bennett et~al.}(2003)}]{wmap1}
\bibinfo{author}{\bibfnamefont{C.~L.} \bibnamefont{Bennett}}
  \bibnamefont{et~al.}, \bibinfo{journal}{\apj}  (\bibinfo{year}{2003}).

\bibitem[{\citenamefont{Spergel et~al.}(2003)}]{wmap2}
\bibinfo{author}{\bibfnamefont{D.~N.} \bibnamefont{Spergel}}
  \bibnamefont{et~al.}, \bibinfo{journal}{\apj}  (\bibinfo{year}{2003}).

\bibitem[{\citenamefont{{\em et al}}()}]{I03}
\bibinfo{author}{\bibfnamefont{K.}~\bibnamefont{{\em et al}}},
  \emph{\bibinfo{title}{Proceedings of the davis meeting on cosmic inflation}},
  \eprint{astro-ph/0304225}.

\bibitem[{\citenamefont{Bond et~al.}(1998)\citenamefont{Bond, Jaffe, and
  Knox}}]{bjk}
\bibinfo{author}{\bibfnamefont{J.~R.} \bibnamefont{Bond}},
  \bibinfo{author}{\bibfnamefont{A.~H.} \bibnamefont{Jaffe}}, \bibnamefont{and}
  \bibinfo{author}{\bibfnamefont{L.}~\bibnamefont{Knox}},
  \bibinfo{journal}{Phys.\ Rev.\ D} \textbf{\bibinfo{volume}{57}},
  \bibinfo{pages}{2117} (\bibinfo{year}{1998}).

\bibitem[{\citenamefont{{Kosowsky} et~al.}(2002)\citenamefont{{Kosowsky},
  {Milosavljevic}, and {Jimenez}}}]{kmj}
\bibinfo{author}{\bibfnamefont{A.}~\bibnamefont{{Kosowsky}}},
  \bibinfo{author}{\bibfnamefont{M.}~\bibnamefont{{Milosavljevic}}},
  \bibnamefont{and}
  \bibinfo{author}{\bibfnamefont{R.}~\bibnamefont{{Jimenez}}},
  \bibinfo{journal}{\prd} \textbf{\bibinfo{volume}{66}}, \bibinfo{pages}{63007}
  (\bibinfo{year}{2002}).

\bibitem[{\citenamefont{{Hu}}(2003)}]{hu02}
\bibinfo{author}{\bibfnamefont{W.}~\bibnamefont{{Hu}}},
  \bibinfo{journal}{Annals of Physics} \textbf{\bibinfo{volume}{303}},
  \bibinfo{pages}{203} (\bibinfo{year}{2003}).

\bibitem[{\citenamefont{Seljak}(2002)}]{uros}
\bibinfo{author}{\bibfnamefont{U.}~\bibnamefont{Seljak}},
  \bibinfo{howpublished}{Private communication} (\bibinfo{year}{2002}).

\end{thebibliography}

\end{document}